\documentclass[10pt, letterpaper, onecolumn]{emulateapj}

\def\Msun{M_\odot}
\def\Mjup{M_J}
\def\microas{\mu{\rm as}}

\begin{document}

\title {Limits to Tertiary Astrometric Companions in Binary Systems}

\author{
Matthew W.~Muterspaugh\altaffilmark{1}, 
Benjamin F.~Lane\altaffilmark{2}, 
S.~R.~Kulkarni\altaffilmark{3},
Bernard F.~Burke\altaffilmark{2},
M.~M.~Colavita\altaffilmark{4},
M.~Shao\altaffilmark{4}}

\altaffiltext{1}{Department of Geological and Planetary Sciences, California 
Institute of Technology, Pasadena, CA 91125}
\altaffiltext{2}{MIT Kavli Institute for Astrophysics and Space Research, 
MIT Department of Physics, 70 Vassar Street, Cambridge, MA 02139}
\altaffiltext{3}{Division of Physics, Mathematics and Astronomy, 105-24, 
California Institute of Technology, Pasadena, CA 91125}
\altaffiltext{4}{Jet Propulsion Laboratory, California Institute of 
Technology, 4800 Oak Grove Dr., Pasadena, CA 91109}

\begin{abstract}
The Palomar High-precision Astrometric Search for Exoplanet Systems (PHASES) 
has monitored 37 sub-arcsecond binary systems to determine whether their 
Keplerian orbits are perturbed by faint astrometric companions to either 
star.  Software has been developed to evaluate the regions in a companion 
mass-period phase space in which the PHASES observations can exclude the 
possibility of face-on orbit perturbations.  We present results for 8 systems 
for which astrometric companions with masses as small as those of giant 
planets can be excluded.
\end{abstract}


\section{Introduction}

A technique has been developed to obtain high precision 
(10-20 $\microas$) astrometry of close stellar pairs \citep[separation less
than one arcsecond;][]{LaneMute2004a} using long-baseline
infrared interferometry at the Palomar Testbed Interferometer 
\citep[PTI;][]{col99}.  
These observations provide precision visual orbits of the binaries and 
allow detection of tertiary components orbiting either the primary or 
secondary due to the reflex motion of the subsystem center-of-light.  

These measurements were made at PTI as part of the 
Palomar High-precision Astrometric Search for Exoplanet Systems
(PHASES) program. PTI is located on Palomar Mountain near San Diego,
CA. It was developed by the Jet Propulsion Laboratory,
California Institute of Technology for NASA, as a testbed for
interferometric techniques applicable to the Keck Interferometer and
other missions such as the Space Interferometry Mission (SIM).  It
operates in the J ($1.2 \mu{\rm m}$), H ($1.6 \mu{\rm m}$), and K
($2.2 \mu{\rm m}$) bands, and combines starlight from two out of three
available 40-cm apertures. The apertures form a triangle with one 110
and two 87 meter baselines.  PHASES observations use two of the three 
available baselines at PTI: the NS (110 meter) and SW (87 meter) baselines.

PTI has previously contributed to exoplanet 
search efforts through high spatial resolution visibility ``imaging'' of 
the 51 Pegasi system\citep{boden1998}, which placed limits on the 
luminosity of its companion and support the conclusion that it is substellar, 
and likely planetary, in nature.  Here the initial detection limits of a 
different effort at PTI are presented, this time based on astrometric 
measurements of binary systems.

\section{The PHASES Exoplanet Search}

PHASES measures the separation vectors of bright binaries that are 
not resolved by PTI's telescopes but are over-resolved by the 
interferometer itself.  The high spatial resolution ($\sim 4$ mas) of the 
interferometer, coupled with an hour of observations, enables extremely 
precise measurement of the binary separation.  Phase-referencing is used 
to maintain coherence so that the full resolving power of the 
interferometer is applied to the astrometric measurement.
The current limiting magnitude in phase-referencing mode is $K = 4.5$ for 
equal magnitude binaries.  Work is being done to improve the sensitivity by 
introducing a variation in which the phase-referencing fringe tracker operates 
at half the current speed, to a 50 Hz duty cycle.
The PHASES program has successfully obtained at least one night of data on 
37 binaries.  As of April 2006, observations have been attempted on 209 
nights, resulting in 688 astrometric measurements.

If only the primary and secondary stars are present in the system, one 
expects these separation vectors to evolve according to a Keplerian model.  
On the other hand, if additional (faint) components are 
present, their presence will cause perturbations to this orbit.  Distant 
companions simultaneously orbiting both visible components will cause only 
very small perturbations to the observed separation vector (the differential 
gravitational pull is small); the PHASES observations are not sensitive to 
these ``circumbinary'' planets \citep[also called ``P-type'' or planetary-type 
planets, ][]{Dvorak1982}.  However, companions orbiting just one star of the 
binary can cause noticeable perturbations to the separation vectors.  This 
configuration is similar to that outlined by the Nemesis theory 
\citep{Nemesis}, which postulated the Sun has a stellar companion orbiting at 
a distance far from the planets.  PHASES is a search for these 
``Nemesis-type'' planetary systems (also called ``S-type'' or satellite-type 
planets).

The primary goal of the PHASES program is to find and characterize 
giant planets in close binary systems.  In this case, ``close'' binary is 
defined as those with semimajor axis $a \lesssim 50 \rm{AU}$ in which one 
might expect binary dynamics to play a major role in system formation and 
evolution \citep[see, for example, ][]{PfahlMute2006}.  The existence of such 
systems poses a strong challenge for models of giant planet formation.  While 
it is possible each of the two processes currently favored---core accretion 
\citep{Liss1993} and gravitational instability \citep{Boss2000}---contribute 
to giant planet formation around single stars and wide binaries, simulations 
show both schemes have obstacles when a second star orbits so closely that it 
interacts with the planet-forming circumstellar disk \citep{Nelson2000}, 
truncating it in size and heating it.  In five exoplanet hosting 
binaries---HD 188753 \citep{Konacki2005}, $\gamma$ Cephei \citep{Hatzes2003}, 
GJ 86 \citep{Queloz2000}, HD 41004 \citep{Zuc2004}, and 
HD 196885 \citep{Chauvin2006}---the secondary star would have truncated the 
disks to less than 7 AU (for systems in which the binary orbits are not fully 
constrained, moderate values for the eccentricities are assumed).  
It is possible that some of these planets formed around single stars or in 
wide binaries and reached their current configurations via dynamical 
interactions in the short-lived star clusters in which they formed 
\citep{Pfahl2005, Portegies_Zwart_2005}, though this post-formation mechanism 
appears to be too infrequent to explain the number found.  

These systems have been identified with the radial velocity (RV) method.  
PHASES employs astrometry, from which the companion mass can be identified 
without the ambiguity of the orbital inclination.  Furthermore, the relative 
orientations of the binary and planetary orbits can be determined in order 
that system dynamics can be studied.  This effort and others specifically 
targeting close binaries will better determine the frequency of planets in 
binaries; if large, this will be strong motivation for a revolution in the 
theory of giant planet formation.

\section{Systematic Errors}

When fit to a low-order polynomial or two-body (single) Keplerian model, the 
current PHASES measurements show night-to-night scatter in excess of that 
determined by evaluating the scatter within a given night.  The timescales of 
this excess scatter can be evaluated by dividing observations from within a 
night into subsets by time and analyzing these subsets individually.  It is 
found that within a given night the subsets agree at the level determined by 
the formal uncertainties obtained from the standard PHASES data reduction 
pipeline.  Because the subsets represent observations separated by of order 
an hour, it is concluded that the PHASES uncertainties are consistent within 
a night; the variations occur on night-to-night timescales.

This excess scatter may be due to either the presence of additional companions 
or due to systematic errors that occur on timescales of days or 
longer.  In order to determine limits to tertiary companions, it 
is assumed that any systematic errors result in true night-to-night 
measurement uncertainties that are related to the formal uncertainties 
by a multiplicative factor.  Depending on the nature of potential 
systematics, this factor may be different for each individual system 
(potentially depending on contrast ratios, separations, or star color 
differences); it is assumed that this factor is a constant for any given 
system over the course of the measurements.  An alternative approach would be 
to postulate a noise floor; this is not evaluated in the current 
investigation.  It is assumed the uncertainties are distributed in a 
Gaussian manner; see section \ref{delEquResiduals}.

\begin{figure}[tbh]
\epsscale{0.6}
\plotone{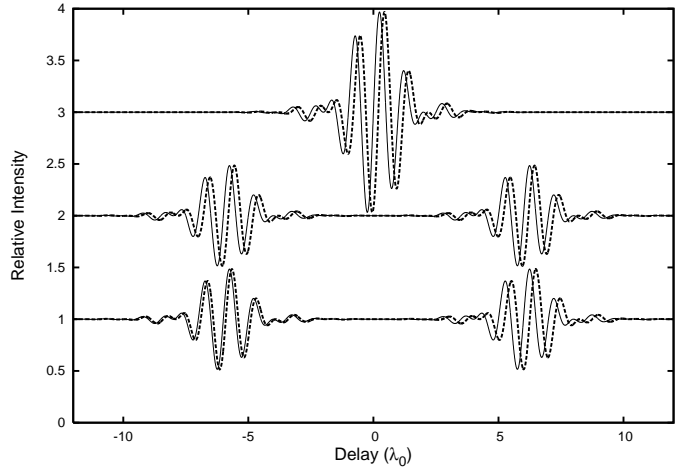}
   \caption[Differential Dispersion] 
{ \label{fig:diff_dispersion}
Schematic of the shift in fringe positions due to dispersion 
(the effect has been exaggerated for clarity).  The vacuum 
(no dispersion) interferograms are plotted with solid lines; 
those dispersed by air with dotted lines.\\
(top) Dispersion shifts the point of zero optical path difference 
for a star, due to different amounts of air path in each 
arm of the interferometer (the effective optical path difference measured 
as if in vacuum).\\
(middle) The dispersion shifts for stars of equal colors are equal and 
cancels; the measured separation is the same.\\
(bottom) Stars of unequal colors are shifted by slightly different amounts 
by dispersion, and the resulting measured separation is different.\\
Not shown are the shape distortions to interferograms.  In a differential 
measurement, these cancel to first order and are insignificant at the 
precisions of PHASES measurements.
}
\end{figure}

\begin{figure}[tbh]
\epsscale{0.6}
\plotone{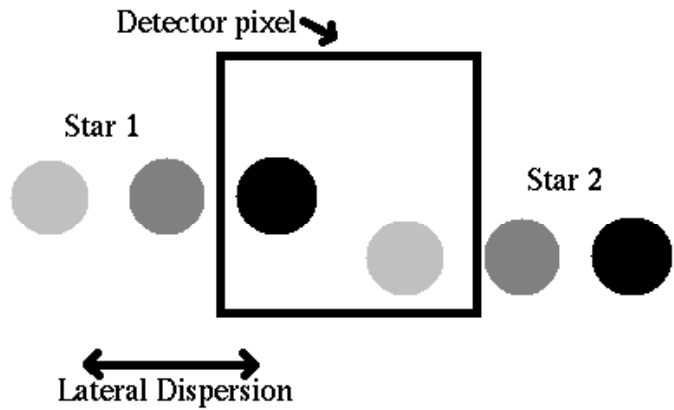}
   \caption[Lateral Dispersion] 
{ \label{fig:lat_dispersion}
Schematic of the shift in fringe positions due to coupling of 
longitudinal and lateral dispersion.  Due to the finite distance between 
stars in the image plane, the detector will sample each star differently.  
If the stars' spectra are dispersed in the image plane, a given detector pixel 
will sample different colors for each star.  This is normally not a problem, 
as the astrometric measurement is not derived from centroiding the stars on 
the detector, but rather from the locations of the movable mirrors at which 
fringes appear on the detector from either one star or the other (each 
star image is the overlap of images from two telescope).  
If there is additionally longitudinal dispersion, the delay location 
at which fringes form will be color dependent.
}
\end{figure}

Potential sources of systematic instrumental errors have been identified and 
efforts are being made to correct them for the next stage of PHASES 
observations (during the 2006-2008 seasons).  The first source is due to 
color dispersion within the interferometer, which manifests itself in two 
manners.  The path compensation for the geometric delay at PTI has been 
done with movable mirrors in air, which has a wavelength-dependent index 
of refraction.  The fringe packets of astrophysical sources are 
dispersed by an amount that depends on the difference in air paths 
between arms of the interferometer; this changes the shape and overall 
location of the fringe packets.  If the two stars are of differing colors, 
each will be dispersed by a slightly different amount, and their 
apparent separation will be biased; see the schematic in figure 
\ref{fig:diff_dispersion}.  Additionally, if there is dispersion 
in a direction perpendicular (lateral) to the light beam (due to 
optics acting as prisms; atmospheric differential chromatic refraction 
also introduces lateral dispersion, though this amount is small) or 
diffraction, the color of the star's light falling on the image plane detector 
pixel can vary with sky position.  In the presence of longitudinal dispersion, 
the dependence of these color shifts on location within the pixel can 
also introduce astrometric systematics; see figure \ref{fig:lat_dispersion}.  
A longitudinal dispersion compensator has been commissioned for PTI which 
should eliminate both systematics.  A 
second source of potential error is drift in the ``astrometric baseline'' due 
to drifts in optical alignment which may result in variable pupil sampling at 
the interferometer apertures.  This is remedied by introduction of an 
automatic alignment system.  These engineering improvements will initiate a 
second stage of the PHASES program, in which it is anticipated the 
observational precisions will be improved by a factor of $\sim 3$.

\section{Detection Limits}

It is anticipated that at least a portion of the excess noise is the result of 
systematic errors, though in some cases the presence of an 
astrometric companion also contributes.  An approach to differentiate between 
these contributions is developed by recognizing that an astrometric 
perturbation will appear as a Keplerian wobble, whereas instrumental terms 
will be random; the Keplerian signature becomes a constraint on the nature of 
the excess scatter.  An algorithm for determining the confidence levels of 
such a signal in the presence of instrumental scatter is described below.

\subsection{Exclusion Criteria}

The hypothesis is that model $\mathcal{A}$ correctly represents the 
observed system; one desires to determine the $\chi^2$ goodness-of-fit 
threshold at which a different model, $\mathcal{B}$, is less consistent 
with the data than $\mathcal{A}$ \citep[for a discussion of 
$\chi^2$ as a likelihood estimator for data with Gaussian 
uncertainties, see for example][]{press92} .  In the present case, model 
$\mathcal{A}$ is the simpler single-Keplerian orbit (with seven 
independent parameters allowed to vary freely, $P_1$, $e_1$, 
$i_1$, $\omega_1$, $T_{o, 1}$, $\Omega_1$, and $a_1$ ), whereas model 
$\mathcal{B}$ is the double-Keplerian model with additional parameters.  In 
the case of a second ``perturbation'' orbit being limited to 
face-on circular configurations, three new free parameters will be 
introduced: period $P_2$, center-of-light semimajor axis $a_2$, and 
one member of the set of the angle of periastron passage $\omega_2$, 
epoch $T_{o, 2}$, and longitude of the ascending node $\Omega_2$, 
which are degenerate; constant parameters are the other two of the 
previous set, the inclination $i_2$, which can either take 
value zero or 180 degrees (though nothing in between), and the 
eccentricity $e_2=0$.  Extension to inclined orbits is discussed in \S 
\ref{sec::inclined}.  A phase space can be constructed over these additional 
free parameters ($P_2$, $a_2$, and $\Omega_2$) to evaluate which combinations 
result in improved or worse fits.  In practice, this phase space is collapsed 
in the $\Omega_2$ dimension to the best-fit value, and $a_2$ is converted to 
tertiary companion mass.

$Q_{\mathcal{A}}$ and $Q_{\mathcal{B}}$ are defined as the number 
of free parameters for models $\mathcal{A}$ and $\mathcal{B}$, 
respectively (here, $Q_{\mathcal{A}} < Q_{\mathcal{B}}$), and there 
are $D$ independent measurements being considered (for astrometry, 
$D$ is twice the number of measures, as each are two dimensional).  
Thus, the numbers of degrees of freedom are 
$Z_{\mathcal{A}} = D - Q_{\mathcal{A}}$ and 
$Z_{\mathcal{B}} = D - Q_{\mathcal{B}}$.

In $\chi^2$ fitting, the best-fit model is expected to have a 
goodness-of-fit $\chi^2$ equal to the number of degrees of freedom.  In 
the hypothesis that model $\mathcal{A}$ accurately describes the data, 
but that the data have excessive noise not included in their formal 
uncertainties, the noise excess factor is given by 
$\chi^2_{\mathcal{A}}/Z_{\mathcal{A}}$.  Reweighting the formal uncertainties 
by the square root of this factor will result in a recomputed value of 
$\chi^2_{\mathcal{A}, {\rm mod}} = Z_{\mathcal{A}}$.  The hypothesis thus 
assumes that this new set of uncertainties correctly represents 
the scatter in the observations as a random process.

When introducing additional free parameters to a model, it is expected that 
the fit $\chi^2$ will decrease.  This does not mean that the model fits 
better, rather this must be placed in terms of the reduced 
$\chi^2_r = \chi^2/Z$.  If the two models represent the data equally well, 
one expects
\begin{displaymath}
\chi^2_{\mathcal{A}}/Z_{\mathcal{A}} = \chi^2_{\mathcal{B}}/Z_{\mathcal{B}}.
\end{displaymath}

If this relationship does not hold and $\mathcal{B}$ is has the larger 
reduced $\chi^2_r$, $\mathcal{B}$ is excluded at the $N\sigma$ level 
for the value of $N$ that satisfies
\begin{displaymath}
\chi^2_{\mathcal{A}}/Z_{\mathcal{A}} = \chi^2_{\mathcal{B}}/ \left( Z_{\mathcal{B}} + \delta_{Q_{\mathcal{B}}, N} \right)
\end{displaymath}
where $\delta_{Q, N}$ is the value by which a properly normalized $\chi^2$ 
metric must be increased to find the $N\sigma$ confidence region, given a 
model with $Q$ free parameters.  In this case, it is the interval computed 
for model $\mathcal{B}$, which by assumption less accurately describes the 
data.

Rearrangement of terms indicates that model $\mathcal{B}$ is excluded 
at or beyond the $N\sigma$ level when 
\begin{equation}\label{eq:exclusion}
\chi^2_{\mathcal{B}} \ge \frac{Z_{\mathcal{B}} + \delta_{Q_{\mathcal{B}}, N}}{Z_{\mathcal{A}}} \chi^2_{\mathcal{A}}.
\end{equation}

\subsection{Astrometric Detection Criteria}

In this hypothesis, we accept $\mathcal{B}$ as the correct model, and look 
for the $\chi^2_{\mathcal{A}}$ level for which model $\mathcal{A}$ is 
excluded.  Thus, we simply replace the roles of the models in equation 
\ref{eq:exclusion}:
\begin{displaymath}
\chi^2_{\mathcal{A}} \ge \frac{Z_{\mathcal{A}} + \delta_{Q_{\mathcal{A}}, N}}{Z_{\mathcal{B}}} \chi^2_{\mathcal{B}}.
\end{displaymath}

Because it is model $\mathcal{B}$ for which a phase space grid is 
constructed (over the extra free parameters in this more complex model), 
it is useful to invert this expression such that it is 
expressed as the contours of $\chi^2_{\mathcal{B}}$ for which model 
$\mathcal{A}$ 
is rejected:
\begin{equation}\label{eq:detection}
\chi^2_{\mathcal{B}} \le \frac{Z_{\mathcal{B}}}{Z_{\mathcal{A}} + \delta_{Q_{\mathcal{A}}, N}}\chi^2_{\mathcal{A}} .
\end{equation}

Table \ref{tab:deltaChiVals} lists values of $\delta_{Q, N}$ used in the 
following sections.
The values of $\delta_{Q, N}$ can be found by iteratively solving the 
equation
$$
P \left( \frac{Q}{2}, \frac{\delta_{Q, N}}{2} \right) = P \left( \frac{1}{2}, \frac{N^2}{2} \right)
$$ 
where 
$$
P(a, x) = \frac{\int_0^x e^{-t}t^{a-1} {\rm d}t}{\int_0^\infty e^{-t}t^{a-1} {\rm d}t}
$$ 
is the standard incomplete gamma function, equal to 
the probability of a $\chi^2$ distribution.  In practice, it 
is often numerically better to equate $Q(a, x) = 1 - P(a, x)$.  Thus, 
the $1\sigma$ contour has probability 0.683, and $3\sigma$ has probability 
0.9973.

One must be careful how the exclusion and detection confidence levels are 
interpreted.  This is particularly true in the case of detections; one must 
not forget the effect of the data sampling function or the possibility that 
a third model better explains the data.  To claim a true astrophysical 
``detection'', one must also satisfy the criteria that the perturbation orbit 
is well constrained.  The exclusion criteria are more robust to such 
possibilities.

\begin{deluxetable}{lll}
\tablecolumns{3}
\tablewidth{0pc} 
\tablecaption{Values of $\delta_{Q, N}$\label{tab:deltaChiVals}}
\tablehead{\colhead{$Q$} & \colhead{$N$} & \colhead{$\delta_{Q, N}$}}
\startdata 
\multicolumn{3}{c}{Model $\mathcal{A}$}\\
5  & 1  &   5.89 \\
5  & 3  &  18.21 \\
6  & 1  &   7.04 \\
6  & 3  &  20.06 \\
6  & 10 & 120.14 \\
7  & 1  &   8.18 \\
7  & 3  &  21.85 \\
7  & 10 & 123.37 \\
7  & 20 & 430.93 \\
7  & 30 & 936.61 \\
\multicolumn{3}{c}{Model $\mathcal{B}$}\\
8  & 1  &   9.30 \\
8  & 3  &  23.57 \\
9  & 1  &  10.42 \\
9  & 3  &  25.26 \\
10 & 1  &  11.54 \\
10 & 3  &  26.90 \\
\enddata
\end{deluxetable}

\subsection{Extension to Inclined Orbits}\label{sec::inclined}

In order for the exclusion limits to be more broadly useful, it is important 
to allow for the possibility of inclined orbits for the tertiary companions.  
The present numerical analysis is limited to face-on orbits only due to 
computational limitations; including a search over a range of orbital 
inclinations introduces two new freely varying parameters, increasing the 
computational time required by roughly two orders of magnitude.  
Alternatively, a procedure can be developed to relate the face-on limits to 
those for inclined systems.

\begin{figure}[tbh]
\epsscale{0.4}
\plotone{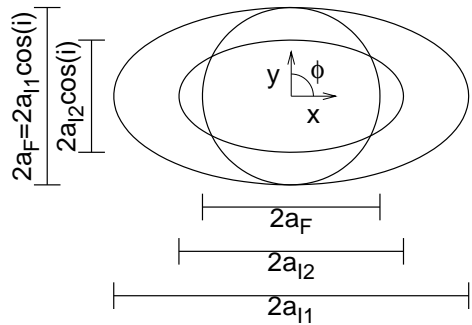}
   \caption[Inclined Orbits] 
{ \label{fig::inclined}
Inclined tertiary component (perturbation) orbits (I1 and I2) 
that fit observations equally well as a face-on orbit model (F).  
Model $I1$ corresponds to the case of measurements with uncertainties 
in dimension $x$ much larger than $y$; model $I2$ corresponds to 
those with equal circular uncertainty ellipses.
}
\end{figure}

For a given face-on system of semimajor axis $a_F$, one can determine the 
size $a_I$ of the inclined orbit model which results in the same $\chi^2$ 
metric; see figure \ref{fig::inclined}.  Two cases are evaluated:  first, when 
the observational uncertainty ellipses are much larger in one dimension than 
the other (case for model $I1$), and second, when the uncertainty ellipses are 
circular (case for model $I2$).  These 
extreme cases bound the values appropriate for PHASES data.  In both cases, 
it is assumed that no perturbation exists and the ``wide'' binary orbit has 
been removed; the residuals are centered at $(x, y) = (0, 0)$.  The $\chi^2$ 
sum is divided into two terms; one summing along the axis parallel to the 
larger dimension of the apparent orbit of the inclined perturbation model 
($x$), and the other along the smaller dimension ($y$):
\begin{equation}\label{eq::splitchi}
\chi^2 = \sum_{i = 1}^{D/2} 
\frac{\left( X \left( \phi \left( t_i \right) \right) - x_i \right)^2}{\sigma_x^2}
+ \sum_{i = 1}^{D/2} 
\frac{\left( Y \left( \phi \left( t_i \right) \right) - y_i \right)^2}{\sigma_y^2}
\end{equation}
where $X(\phi(t)) = a \cos \phi$ and $Y(\phi(t)) = a \cos i \sin \phi$ ($i$ is the 
inclination of the orbit) are the 
perturbation orbit model values at time $t$ 
in each dimension ($\phi = 2\pi t / P + \phi_o$; $\phi_o$ is a phase offset), 
and $\sigma_x^2$ and $\sigma_y^2$ are the respective 
uncertainties for each dimension.  There are no $x-y$ cross terms by choice 
of basis, explained below.  Conditions for which the $\chi_F^2$ of the face-on 
orbit equals $\chi_{I1}^2$ or $\chi_{I2}^2$ of the inclined models are sought.

In the case of non-circular uncertainty ellipses, one works in a basis for 
which the uncertainty ellipse major axis is oriented along the $x$ axis.  It 
is assumed the uncertainty ellipses share a single such basis (their 
orientations 
do not vary) such that a $\chi^2$ minimizing algorithm will align the larger 
dimension of the apparent perturbation orbit with the large uncertainty axis 
(this also assumes there are no external limits placed on the longitude of 
the ascending node).  
For PHASES data, the orientations of the uncertainty ellipses vary slightly 
from night to night (depending on the hour angle of observations); this serves 
only to lessen the resulting conversion factor, and the present derivation is 
for the more extreme case.  In the limit of large uncertainties in the $x$ 
direction, $\mid X(t) \mid \le a_{I1} \ll \sigma_x^2$ and the first term in 
eq.~\ref{eq::splitchi} is unchanged and cancels upon setting 
$\chi_{F}^2 =\chi_{I1}^2$, leaving 
\begin{equation}
\sum_{i = 1}^{D/2} 
\frac{\left( Y_F \left( t_i \right) - y_i \right)^2}{\sigma_y^2} = 
\sum_{i = 1}^{D/2} 
\frac{\left( Y_{I1} \left( t_i \right) - y_i \right)^2}{\sigma_y^2}.
\end{equation}
By geometry, $Y_F(\phi) = a_F \sin \phi$ and 
$Y_{I1}(\phi) = a_{I1} \cos i \sin \phi$.  
By setting $a_F = a_{I1} \cos i$, $\chi_{F}^2 =\chi_{I1}^2$.  
Thus, inclined orbits with semimajor axis $1/\cos i$ times larger will 
fit the observations equally well as a face-on model.

For circular uncertainty ellipses, $\sigma_x^2 = \sigma_y^2 = \sigma^2$
\begin{equation}
\begin{array}{ccc}
\sum_{i = 1}^{D/2} 
\frac{\left( X_F \left( \phi \right( t_i \left) \right) - x_i \right)^2}{\sigma^2} 
& & \\
+ \sum_{i = 1}^{D/2} 
\frac{\left( Y_F \left( \phi \right( t_i \left) \right) - y_i \right)^2}{\sigma^2}
& = & \sum_{i = 1}^{D/2} 
\frac{\left( X_{I1} \left( \phi \right( t_i \left) \right) - x_i \right)^2}{\sigma^2} \\
 & & + \sum_{i = 1}^{D/2} 
\frac{\left( Y_{I2} \left( \phi \right( t_i \left) \right) - y_i \right)^2}{\sigma^2} \\
\sum_{i = 1}^{D/2} \left( a_F \cos \phi \left( t_i \right) - x_i \right)^2
& & \\
+ \sum_{i = 1}^{D/2} \left( a_F \sin \phi \left( t_i \right) - y_i \right)^2
& = & \sum_{i = 1}^{D/2} \left( a_{I2} \cos \phi \left( t_i \right) - x_i \right)^2 \\
 & & + \sum_{i = 1}^{D/2} \left( a_{I2} \cos i \sin \phi \left( t_i \right) - y_i \right)^2.\\
\end{array}
\end{equation}
Assuming the measurements are uniformly distributed in $\phi$, and $x_i$ and $y_i$ have 
statistically similar distributions $p$ which are even functions (such as Gaussians), this relationship becomes
\begin{small}
\begin{equation}
\begin{array}{ccc}
\int_{-\infty}^{\infty} \int_0^{2\pi} \left( a_F \cos \phi - x \right)^2 p(x) {\rm d}\phi {\rm d}x   &   & \\
+ \int_{-\infty}^{\infty} \int_0^{2\pi} \left( a_F \sin \phi - y \right)^2 p(y) {\rm d}\phi {\rm d}y & = & 
\int_{-\infty}^{\infty} \int_0^{2\pi} \left( a_{I2} \cos \phi - x \right)^2 p(x) {\rm d}\phi {\rm d}x\\
 & & + \int_{-\infty}^{\infty} \int_0^{2\pi} \left( a_{I2} \sin \phi \cos i - y \right)^2 p(y) {\rm d}\phi {\rm d}y\\
\int_{-\infty}^{\infty} \int_0^{2\pi} \left( a_F^2 \cos^2 \phi - 2 a_F \cos \phi x + x^2 \right) p(x) {\rm d}\phi {\rm d}x & & \\
\int_{-\infty}^{\infty} \int_0^{2\pi} \left( a_F^2 \sin^2 \phi - 2 a_F \sin \phi y + y^2 \right) p(y) {\rm d}\phi {\rm d}y & = & 
\int_{-\infty}^{\infty} \int_0^{2\pi} \left( a_{I2}^2 \cos^2 \phi - 2 a_{I2} \cos \phi x + x^2 \right) p(x) {\rm d}\phi {\rm d}x\\
 & & + \int_{-\infty}^{\infty} \int_0^{2\pi} \left( a_{I2}^2 \sin^2 \phi \cos^2 i - 2 a_{I2} \sin \phi \cos i y + y^2 \right) p(y) {\rm d}\phi {\rm d}y \\
\int_{-\infty}^{\infty} \int_0^{2\pi} a_F^2 \cos^2 \phi p(x) {\rm d}\phi {\rm d}x & & \\
+ \int_{-\infty}^{\infty} \int_0^{2\pi} a_F^2 \sin^2 \phi p(y) {\rm d}\phi {\rm d}y & = & \int_{-\infty}^{\infty} \int_0^{2\pi} a_{I2}^2 \cos^2 \phi p(x) {\rm d}\phi {\rm d}x\\
 & & + \int_{-\infty}^{\infty} \int_0^{2\pi} a_{I2}^2 \sin^2 \phi \cos^2 i p(y) {\rm d}\phi\\
\end{array}
\end{equation}
\end{small}
where in the last step, the second terms (linear in $x$ and $y$) integrate to 
zero because $p$ is even, and the third terms cancel on the right and left sides of 
the equation.  Finally, the remaining $x$ and $y$ integrals are common to all terms 
and cancel, and the $\phi$ integrals evaluate to identical values for all terms, leaving
\begin{equation}
2 a_F^2 = a_{I2}^2 \left( 1 + \cos^2 i \right).
\end{equation}
Thus, by setting $a_F = a_{I2} \sqrt{\left( 1 + \cos^2 i \right)/2}$, $\chi_F^2 = \chi_{I2}^2$.

Because orbital size is proportional to companion mass, these multiplicative 
factors can be directly applied to the exclusion limits for astrometric 
companions.  To convert between the results for face-on and those including 
inclined models, one need multiply only by a factor between 
$\sim \sqrt{2/(1+\cos^2 i)}$ and $\sim 1/\cos i$.  The exact value of 
this factor depends on the distributions of observations in time and 
orientations and aspect ratios of the uncertainty ellipses.

\section{$\delta$ Equulei}

PHASES observations of the nearby binary system $\delta$ Equulei 
(7 Equ, HR 8123, HD 202275, ADS 14773), including an updated 
visual orbit, have been presented by \cite{Mut05_delequ}.  The 27 
observations reported there have been combined with 11 new measurements from 
the 2006 observing season to search for evidence of astrometric companions 
around either star.  Figure \ref{fig:202275_phase_space} shows the regions 
in companion mass-period phase space for which the measurements are 
inconsistent with a perturbation caused by a hypothetical companion in a 
face-on circular orbit.  Companions with masses greater than the lines plotted 
are inconsistent with the PHASES observations; the exclusion regions 
are those above the lines.

\subsection{Application of Detection Limits Procedure}

Implementation of the criteria described by equations \ref{eq:exclusion} and 
\ref{eq:detection} can be demonstrated on this system.  The binary orbit is 
short enough that it is well constrained by the PHASES observations 
alone ($P_1 < 6$ years); the elements of the visual orbit are not strongly 
covariant.  Thus, all seven parameters of a visual orbit 
($P_1, e_1, i_1, \omega_1, T_{o, 1}, \Omega_1, a_1$) are 
allowed to be varied in Keplerian model 
($Q_{\mathcal{A}} = 7$).  The resulting $\chi^2_{\mathcal{A}}$ is 
1090.83.  The 38 PHASES measurements are each two-dimensional, the 
expected $\chi^2_{\mathcal{A}}$ is thus 
$ Z_{\mathcal{A}} = 2 \times 38 - 7 = 69 $; under the hypothesis that 
the single-Keplerian model is correct, $\chi^2_{\mathcal{A}}$ is in excess 
by a multiplicative factor of 1090.83/69 = 15.81.  The measurement 
uncertainties themselves are too small by a factor of $\sqrt{15.81} = 3.98$, 
and the median minor-axis formal uncertainty of $7.7\microas$ is corrected to 
$30.6 \microas$.

Model $\mathcal{B}$ is a double-Keplerian orbit given by superposition as 
\begin{eqnarray}\label{doubleKeplerianCOL}
\overrightarrow{y_{\rm{obs}}} &=& \overrightarrow{r_{\rm{A-B}}}
\left( P_1, e_1, i_1, \omega_1, T_{o, 1}, \Omega_1, a_1 \right) + \nonumber \\
& & \overrightarrow{r_{\rm{Ba-Bb, C.O.L.}}}
\left( P_2, e_2, i_2, \omega_2, T_{o, 2}, \Omega_2, a_2 \right).
\end{eqnarray}
Here, the A-B orbit is the slow, ``wide'' orbit and the Ba-Bb orbit is 
the ``narrow'' perturbation orbit center-of-light motion.  
When converting $a_2$ to $M_{\rm {Bb}}$, assumed values for 
$M_{\rm Ba} = 1.19 \Msun$ and the system distance $d = 18.386$ pc are used, 
and it is assumed that component Bb is faint (i.e.~the Ba-Bb 
center-of-light is located at Ba, and $a_2 = a_{\rm {Ba}}$, corresponding to 
just the reflex motion of the star).  In $\delta$ 
Equ, both stellar components A and B mass roughly $1.19 \Msun$; the 
model is symmetric for companions to A or B and the derived limits on 
companion masses are identical for the two---one search eliminates companions 
around both stars, in this case with nearly equal mass limits.

To evaluate whether model $\mathcal{B}$ (equation 
\ref{doubleKeplerianCOL}) better fits the observations, the $\chi^2$ 
metric is used to evaluate the goodness-of-fit 
at each point in a grid of hypothetical companion masses and 
orbital periods.  The grid was stepped logarithmically in 
companion mass from $10^{-1}$ to $10^3$ $\Mjup$ with step 
size $\log\left(M_{\rm {Bb}}/\Mjup\right) = 0.1$.  The companion 
period was sampled at values of $P_2 = 2fT/K$, where $f$ is an excess 
factor (here 3) for finer sampling, $T$ is the span of the observations 
($\sim 760$ days), and $K$ is a natural number up to that for which 
$P_2 = 3$ days.  $P_2$ was additionally evaluated at values 
of $K = 1/2, 1/3, ..., 1/9$.

At each grid point, the seven parameters of the wide orbit are seeded with 
those from the best-fit single-Keplerian 
(i.e.~model $\mathcal{A}$); the perturbation orbit's 
parameters are seeded in period and (center-of-light) semimajor axis as 
determined by the phase-space grid.  Additionally, the perturbation 
orbit is seeded in several values of $\Omega_2$ (separated by 40 
degrees starting at 0 and increasing up to 320), and at inclinations 
of either zero or 180 degrees to allow for face-on orbits 
with clockwise or counter-clockwise motion; all combinations of these 
$\Omega_2$ and $i_2$ seedings are explored.  The values of the other 
secondary orbit parameters are seeded at zero.  

The model is allowed to relax from each set of initial seedings 
using a downhill $\chi^2$ fitting algorithm in which the seven 
parameters of the ``wide'' orbit and $\Omega_2$ in the ``narrow'' orbit are 
allowed to vary.  Note that the narrow system inclination is not allowed to 
vary---this restricts the search to including only face-on orbits, and the 
inclination parameter does not add to the count of ``free parameters'' in the 
model.  While the period and semimajor axis (and thus mass) of the 
perturbation orbit are not allowed to vary in the downhill fitting process, 
these are varied across the grid and are to be counted in the total 
number of model parameters.  Note that with eccentricity held at zero 
(circular orbits) and inclinations restricted to zero and 180 degrees, 
$\Omega_2$ is degenerate with $T_{o, 2}$ and $\omega_2$, thus these need not 
be varied.  The total number of degrees of freedom in the double-Keplerian 
model is thus $Q_{\mathcal{B}} = 10$; seven are from the wide orbit, plus the 
period $P_2$, orbit phase $\Omega_2$, and semimajor axis $a_2$ 
(or companion mass) of the perturbing orbit.  From this, 
$Z_{\mathcal{B}} = Z_{\mathcal{A}} - 3 = 66$. 

\begin{figure}[tbh]
\epsscale{0.4}
\plotone{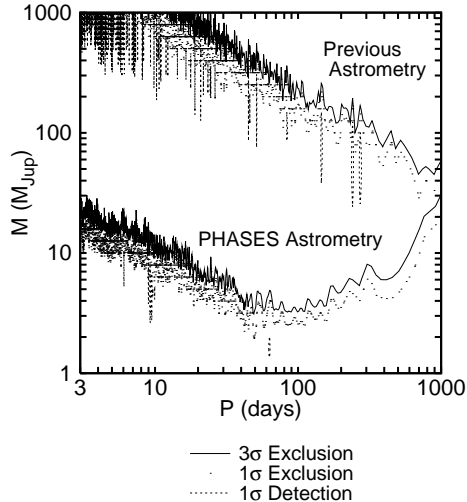}
\caption[$\delta$ Equulei Mass-Period Companion Phase Space] 
{ \label{fig:202275_phase_space}
The $\delta$ Equulei Mass-Period companion phase space.  
Companions in the regions above the plotted exclusion curves with face-on 
orbits are not consistent with the observations.  Companions as small as 
two Jupiter masses can be ruled out by PHASES observations.  The 
most significant detections are less than $2\sigma$ significant, and are not 
likely astrophysical in origin.}
\end{figure}

The $N\sigma$ exclusion regions are defined as those for which the 
double-Keplerian model $\chi^2_{\mathcal{B}}$ is greater than 
$1090.83 \times (66+\delta_{10, N})/69$.
The $N\sigma$ detection regions are calculated from 
equation \ref{eq:detection} as being those for which the 
double-Keplerian $\chi^2_{\mathcal{B}}$ is less than 
$1090.83 \times 66 / (69 + \delta_{7, N})$.
The $1\sigma$ ($\chi^2_{\mathcal{B}} = 1226$) and $3\sigma$ 
($\chi^2_{\mathcal{B}} = 1469$) exclusion contours in mass-period phase 
space are plotted in figure \ref{fig:202275_phase_space}; also shown is 
the $1\sigma$ {\em detection} contours at $\chi^2_{\mathcal{B}} = 933$; 
the smallest value of $\chi^2_{\mathcal{B}}$ is 883, well above the 
$2\sigma$ detection criteria of $\chi^2_{\mathcal{B}} = 864$.  
It is concluded these detection regions are results of statistical 
happenstance.

\subsection{Comparison to Previous Work}

The same procedure has been applied to the non-PHASES astrometric 
measurements reviewed by \cite{Mut05_delequ}; see this work for a discussion 
of the relative weightings of these measurements.  These 170 measurements 
cover a much longer time span (of a century) than the PHASES observations, 
but at lower precisions.  In this case, the value of $T$ used to determine the 
companion periods sampled was chosen to be half the period of the wide orbit 
found in the best fit single-Keplerian model ($P_1/2 = T \sim 1042$) days.  
The best fit seven-parameter single-Keplerian model results 
in $\chi^2_{\mathcal{A}}=426.5$ 
and $Z_{\mathcal{A}} = 340 - 7 = 333 $ degrees of freedom.  Thus, 
the $1\sigma$ and $3\sigma$ exclusion level for $\chi^2_{\mathcal{B}}$ of 
the ten-parameter double-Keplerian model are 437.5 and 457.2, respectively.  
For $P_2 < P_1$, no values of $\chi^2_{\mathcal{B}}$ in the grid are less than 
400, which is between the $2\sigma$ and $3\sigma$ detection thresholds (this 
lowest point is near $P_2 \sim 10.25 $ days and $M_{\rm {Bb}} \sim 800 \Mjup$, 
which is strongly disfavored by the PHASES data).

\subsection{Inclined Orbits}

At this time, it is computationally prohibitive to search in additional 
parameters (for example, to add an array of seeds in inclination and allow 
this parameter to also vary, or to search for eccentric orbits).  It is noted 
that there is always an orbit for which astrometry is completely insensitive 
to companions of any mass or period---high eccentricity, edge-on orbits 
pointing directly at the observer (i.e.~periastron passage within the line of 
sight); combination fitting with radial velocity observations lifts this 
problem and can be useful when contemplating universal exclusions of 
perturbing companions.  Evaluating the phase-space grid described above using 
just the PHASES astrometry and a 10 parameter model requires roughly 3 weeks 
on a modern processor; each additional parameter is expected to increase this 
by roughly an order of magnitude.  A need is identified for an alternative 
approach than the ``brute-force'' algorithm described here, in combination 
with an evaluation of its performance and reliability in comparison to the 
method described here.

To test the analytical conversion factor between detection limits for face-on 
and inclined systems of \S \ref{sec::inclined}, the PHASES observations of 
$\delta$ Equulei were reanalyzed allowing inclinations to vary within 45 
degrees of face-on (both prograde and retrograde orbits were allowed) over a 
very limited subset of perturbation periodicities.  The 
predicted conversion factor is between $\sqrt{4/3}$ and $\sqrt{2}$; in 
practice, a factor of $\sim 1.35-1.4$ is appropriate for the PHASES 
observations.

\subsection{Fit Residuals}\label{delEquResiduals}

\begin{figure}[tbh]
\epsscale{0.6}
\plotone{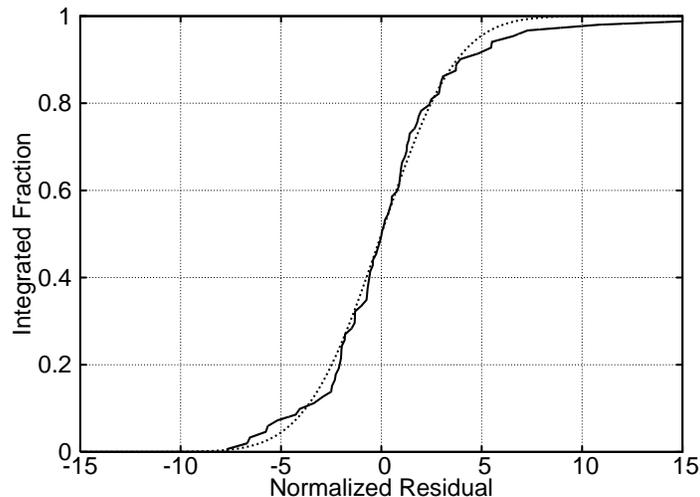}
\caption[$\delta$ Equulei Fit Residuals Distribution]
{ \label{fig:residuals}
The distribution of normalized residuals for a 2-body Keplerian fit to 
$\delta$ Equulei.  Also shown is the theoretical distribution for Gaussian 
noise.}
\end{figure}

Figure \ref{fig:residuals} shows the continuous (integrated) distribution 
function of the residuals to the 2-body Keplerian fit for the $\delta$ Equulei 
system.  For each measurement, the residuals were measured along directions 
parallel to the formal uncertainty error ellipse major and minor axis in 
order that the uncertainties are not covariant.  Each were then normalized by 
that measurement's formal uncertainty estimate.  The distribution of this 
normalized set of residuals can now be considered to determine whether a 
multiplicative scale factor is an appropriate assumption for handling the 
observed excess scatter, in which case the residuals should have a Gaussian 
distribution.  The continuous distribution function is fit to the integral 
of a Gaussian:
\begin{equation}
\frac{1+{\rm erf}(x/\sqrt{2}a)}{2}
\end{equation}
where $a$ is a measure of the excess scatter over the formal uncertainties 
(the best fit value for $a$ is $2.94\pm0.01$).  
A one-sided Kolmogorov-Smirnov test 
\cite[see for example section 14.3 of ][]{press92} shows the residual 
distribution agrees with Gaussian statistics with $95\%$ likelihood.  
The distributions match well and the assumption of Gaussian errors 
and multiplicative uncertainty scale factors is valid.  Treated as separate 
sets, the scale factors for the major-axis and minor-axis residuals are 
similar.

\section{Known Triple Stars}

This procedure is tested on two binaries that each are known to host 
tertiary companions, $\kappa$ Pegasi and V819 Herculis.  The PHASES 
astrometric orbits of these triples have been previously examined 
\citep[][]{Mut06_kappeg, Mut06_v819her}; the phase space exploration 
algorithm detects these perturbations and their statistical significances 
are now presented.

\subsection{$\kappa$ Pegasi}

PHASES observations of the well-known triple system $\kappa$ 
Pegasi (10 Peg, HR 8315, HD 206901, ADS 15281), from 
which the orbits of both the long and short period subsystems 
were determined, have been presented by \cite{Mut06_kappeg}.  The 
52 PHASES measurements presented there are combined with three 
measurements from 2005 creating a set spanning $\sim 1046$ days.  
In the previous investigation, both a noise floor and multiplicative 
factor were explored for explaining excess systematic noise; the 
present investigation evaluates only the latter case.

A mass-period grid was constructed similar to that in the case of $\delta$ 
Equulei, differing in the following ways:  the $P_2$ oversample factor $f$
is 1 rather than 3, the companion masses explored ranged from 
$10^{0}$ to $10^4$ $\Mjup$, and the step size in companion mass was 
$\log\left(M_{\rm {Bb}}/\Mjup\right) = 0.2$ rather than 0.1.  The 
424 non-PHASES astrometry data tabulated in the previous investigation were 
similarly evaluated over the same grid; because these span a time much 
longer than the wide orbit period, $T$ was set to half the best-fit wide 
system period $P_1$ (for evaluation of the PHASES data, $T$ is the timespan 
they cover).  For both the PHASES and non-PHASES data sets, the wide orbit 
is well-constrained without strong covariances.  Thus, for both cases 
$Q_{\mathcal{A}} = 7$ and $Q_{\mathcal{B}} = 10$.  The companion Bb mass to 
$a_2$ conversion assumed a stellar mass $M_{\rm {Ba}} = 1.662 \Msun$ and 
distance to the system of 34.60 pc, as determined by the previous 
investigation.  Note that this analysis is only to confirm the detection 
of the known 5.97-day period companion Bb; at this time no attempt is 
made to search for fourth components to the system, which may require much 
more complicated modeling than the simple superpositioning of independent 
Keplerians that has been used here.

\begin{figure}[tbh]
\epsscale{0.5}
\plotone{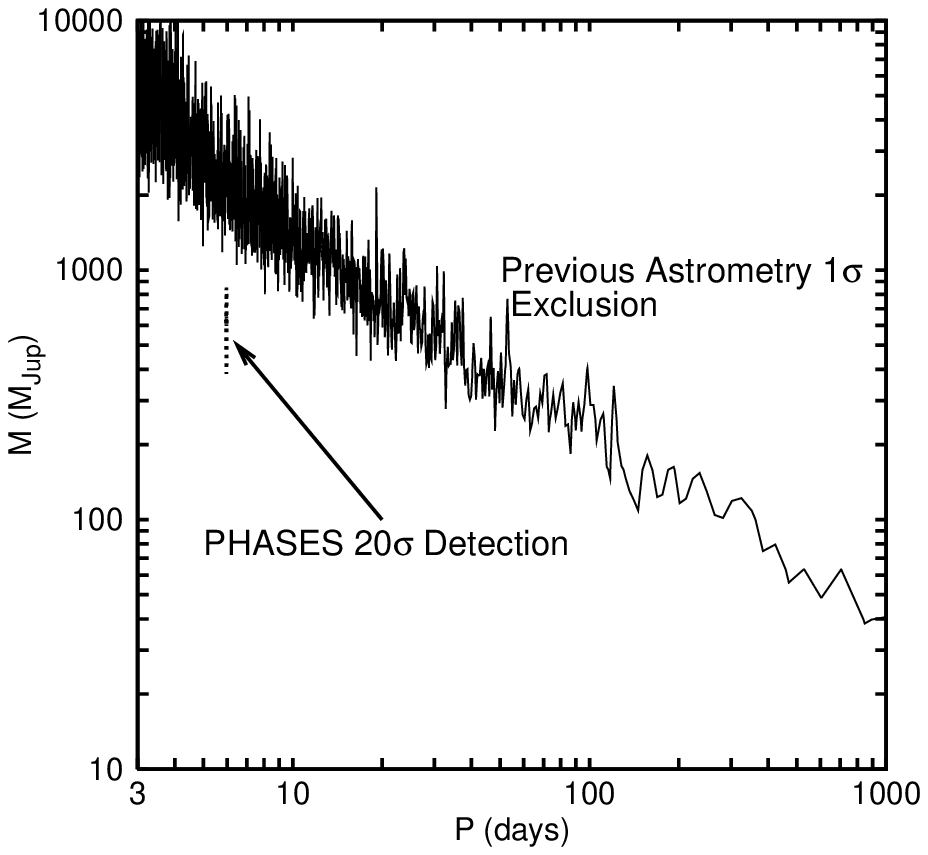}\\
\plotone{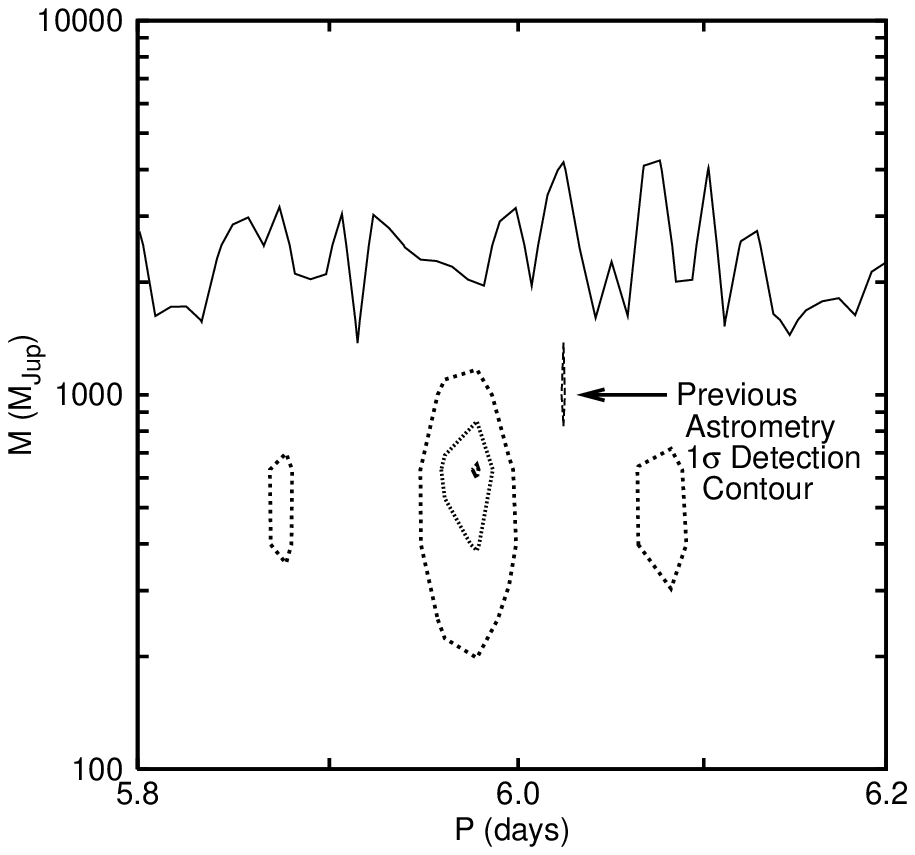}
\caption[$\kappa$ Pegasi Mass-Period Companion Phase Space] 
{ \label{fig:206901_phase_space}
$\kappa$ Pegasi Mass-Period Companion Phase Space.\\
(Top) The $1\sigma$ exclusion contour based on previous differential 
astrometry data, with the $20\sigma$ PHASES detection contour.  Note 
that only at the 5.97 day companion period does the $20\sigma$ PHASES contour 
exist.\\
(Bottom)  The 10-, 20-, and 30-$\sigma$ detection contours for PHASES 
observations 
of the $\kappa$ Pegasi triple star system are plotted for the limiting 
case of face-on Ba-Bb subsystem orbits.  While the Ba-Bb subsystem is 
not in fact face-on, the perturbation is detected with the same periodicity 
as that found by RV observations.  Once this periodicity has been identified, 
the face-on, circular orbit constraint can be removed, and the detection 
becomes $106\sigma$ significant.  Also plotted is the $1\sigma$ exclusion 
contour based on previous differential astrometry data.  
A ``detection'' at the $1\sigma$ level in the previous differential astrometry 
is seen in the plot; however, this is likely due only to statistical 
happenstance.
}
\epsscale{1.0}
\end{figure}

In the PHASES evaluation, the smallest value of $\chi^2_{\mathcal{B}}$ is 
found at $P_2 = 5.978$ days; in this region the step size is 
$\Delta P_2 \sim 0.017$.  This is consistent with the best-fit value of $P_2$ 
of $5.9714971 \pm 1.3 \times 10^{-6}$ days.  The depth of this detection 
corresponds to a detection at the $31.4\sigma$ level.  Conversely, 
the smallest value of $\chi^2_{\mathcal{B}}$ for the non-PHASES data is 
at $P_2 = 19.14$ days with depth not quite reaching the $3\sigma$ detection 
level.  The smallest value with $5.9 < P_2 < 6.1$ days is just barely 
smaller than the $1\sigma$ threshold (near $1.1\sigma$).  Figure 
\ref{fig:206901_phase_space} plots the relevant phase-space contours.

Upon removing the face-on, circular orbit criteria for the narrow orbit 
(seeding a $\chi^2$ downhill fit with the best set of parameters found in 
the phase-space analysis), the fit is considerably improved, 
$Q_{\mathcal{B}} = 14$, and the second orbit improves the fit at 
the $106\sigma$ level.


\subsection{V819 Herculis}

PHASES observations have previously determined the astrometric orbit of the 
short-period subsystem in the V819 Herculis (HR 6469, HD 157482) triple system 
\citep{Mut06_v819her}.  The narrow (Ba-Bb) pair of the V819 Her system is 
oriented edge-on and show eclipses.  In that analysis, 31 measurements were 
reported (of which six were not used in model fitting as they were taken 
during eclipse of the narrow pair); one additional measurement from 
summer 2005 is added to this set.  In the present analysis, measurements 
taken during eclipse are not omitted to simulate a blind search.

When only the PHASES data are fit, the span of the measurements 
($T \sim 476$ days) is much shorter than the period of the wide orbit 
($P_1 \sim 2020$ days).  Thus, $P_1$, $e_1$, and $a_1$ are strongly 
covariant; by fixing just one of these (here, $P_1 = 2019.787$ days) at 
a nominal value supported by non-PHASES observations, these covariances 
are lifted and the fitting algorithm converges much more rapidly.  Because 
that parameter is not optimized for each fit, it does not count to the 
number of free parameters; for the PHASES analysis, $Q_{\mathcal{A}} = 6$ 
and $Q_{\mathcal{B}} = 9$; this is not necessary when the non-PHASES 
observations are included, for which $Q_{\mathcal{A}} = 7$ and 
$Q_{\mathcal{B}} = 10$ as normal.

As in the case of $\kappa$ Pegasi, the mass-period phase space grid was 
modified by setting the $P_2$ oversample factor $f$ to 1 rather 
than 3, the companion masses explored ranged from 
$10^{0}$ to $10^4$ $\Mjup$, and the step size in companion mass was 
$\log\left(M_{\rm {Bb}}/\Mjup\right) = 0.2$ rather than 0.1.  Additionally, 
the smallest value of $P_2$ sought was 1 day rather than 3.  The distance 
is assumed to be 67.96 pc and $M_{\rm {Ba}} = 1.43 \Msun$.  Again, the aim of 
the search is only to confirm detection of the known companion Bb, and no 
attempt is made to find fourth components at this time.

Because the algorithm has been developed to specialize in face-on 
orbits, it is not optimally suited to analysis of the V819 
Her system, for which the short-period perturbation system is observed 
to eclipse.  However, the strength of the algorithm as a ``wobble-detector'' 
is demonstrated by applying it to this system for which it is quite 
non-optimally designed.  The two deepest dips in the $\chi^2$ function 
occur at periodicities of 2.2284 days 
($\Delta P_2 \sim 0.0052$ days; $10.5\sigma$) 
and 1.8056 days ($\Delta P_2 \sim 0.0034$ days; $10.1\sigma$).
These are the only two points at the $10\sigma$ level; the 1.8056 day 
periodicity appears to be only an aliasing; 
the RV and eclipse photometry coincide with the (more significant) 2.2284 day 
periodicity (the best fit period to all data is 
$P_2 = 2.2296337\pm 1.9\times 10^{-6}$ days, in agreement with the 
2.2284 periodicity, to within the analysis step size).  This short-period, 
$110 \microas$ perturbation is readily detected, but is far beyond the 
ability of other astrometric measurements; with 
this periodicity identified, one can better refine the fit while 
allowing other parameters to vary, obtaining the correct orbital 
configuration for the short-period pair and improving the detection 
to the $14.7\sigma$ level.

\begin{figure}[tbh]
\epsscale{0.4}
\plotone{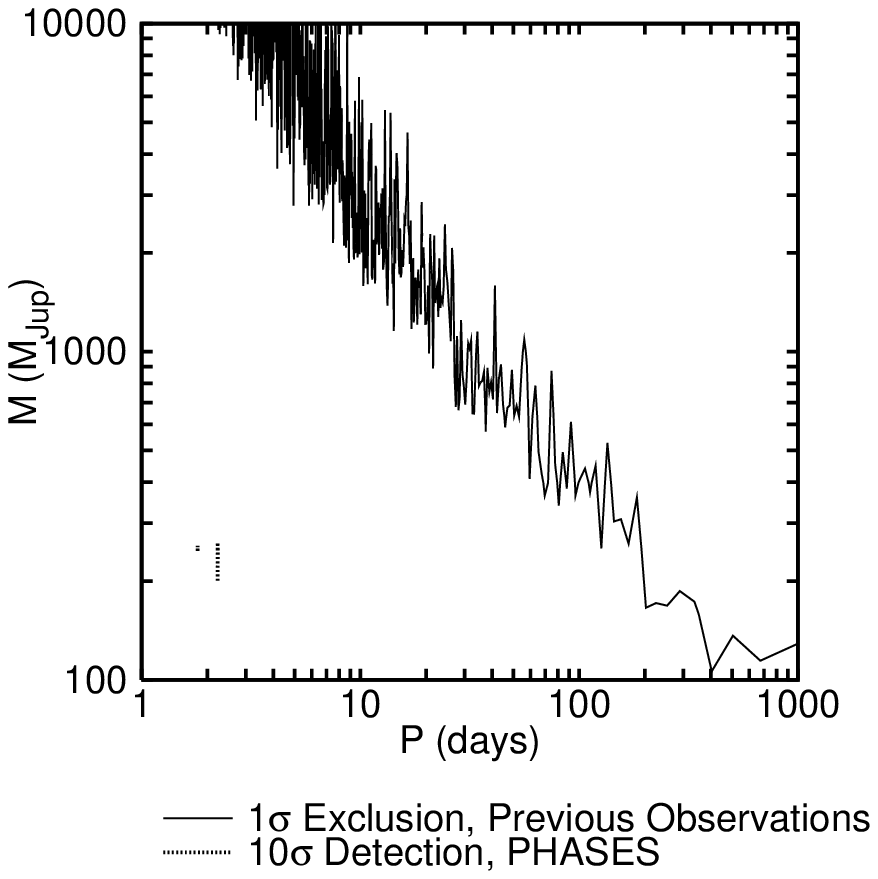}
\caption[V819 Herculis Mass-Period Companion Phase Space] 
{ \label{fig:157482_phase_space}
The V819 Her Mass-Period companion phase space shows a $10\sigma$ significant 
detection in the PHASES data that is far beyond the detection threshold for 
previous measurements.  Although the tertiary companion's orbit is in fact 
edge-on, the face-on code detects the wobble at the correct period.  After 
removing the face-on orbit restraint, the double-Keplerian model is a 
$14.7\sigma$ improvement over the single-Keplerian model.
}
\end{figure}

The 34 speckle interferometry data tabulated in the previous paper are 
evaluated separately (in this case $P_1$ is allowed to vary freely).  As 
shown in figure \ref{fig:157482_phase_space}, the speckle observations are 
approximately two orders of magnitude away from detecting the 
short-period perturbation.  Interestingly, the $110 \microas$ amplitude 
perturbation would correspond to a planetary mass companion if only its 
period were several months rather than several days.  

\section{$\beta$ Coronae Borealis}

$\beta$ Coronae Borealis (``Peculiar Rosette Stone'', 3 CrB, HR 5747, 
HD 137909) is a binary whose primary is of spectral class F0p, part of the 
family of peculiar A stars with strong magnetic fields.  $\beta$ CrB is 
often grouped with $\gamma$ Equulei and $\alpha^2$ CVn as prototypes of 
this class.  Its magnetic field has been extensively via measurements 
of Zeeman line splitting, and the inclination of its rotation axis and 
the angular offset of its magnetic field have been measured.

\cite{Neubauer1944} studied the binary using RV measurements and 
found evidence for a third body with orbiting the primary with 
a period of nearly a year ($P_2 \sim 320$ days).  \cite{Kamper1990} 
obtained updated RV data which did not show 
this perturbation.  However, their analysis 
showed the periodicity in \citeauthor{Neubauer1944}'s data is 
statistically different than one year.  They suggested the perturbation 
was real rather than instrumental, and proposed that this orbit 
was previously nearly face-on, somehow re-orienting itself to be 
perfectly face-on between the two epochs.  \cite{Soder1999} concluded 
from {\em Hipparcos} astrometric observations that such a proposed 
companion cannot exist.

Forty-two PHASES measurements have been collected over a span of $\sim 804$ 
days.  Additionally, 102 non-PHASES differential astrometry measurements 
have been identified (when these are analyzed, $T$ is set to half the best 
fit value of $P_1$ in the 2-body single-Keplerian case).  
As was the case for V819 Her, $P_1$, $e_1$, and $a_1$ are strongly 
covariant when only PHASES data are considered ($T \sim 804$ days versus 
$P_1 \sim 10.27$ yr).  Again, $P_1$ is fixed at a value that is not 
based on the PHASES observations; $P_1 = 10.27$ yr is adopted from 
\cite{Tok1984}, and $Q_{\mathcal{A}} = 6$, $Q_{\mathcal{B}} = 9$ 
(7 and 10 when non-PHASES data are included).  

The mass-period grid is set up in the same manner as for $\delta$ Equulei.  
Based on analysis of {\em Hipparcos} observations by \cite{Soder1999}, the 
distance is assumed to be 34.12 pc ($\pi = 29.31 \pm 0.82$ mas) and the 
component masses are $1.77 \pm 0.16$ and $1.21 \pm 0.11\Msun$.  The 
conversion of tertiary mass to semi-major axis assumes a stellar mass of 
$1.21 \Msun$.  Radial velocity measurements are available only for one 
component of the binary; this prevents one from determining the masses and 
distance from the orbit alone.

\begin{figure}[tbh]
\epsscale{0.4}
\plotone{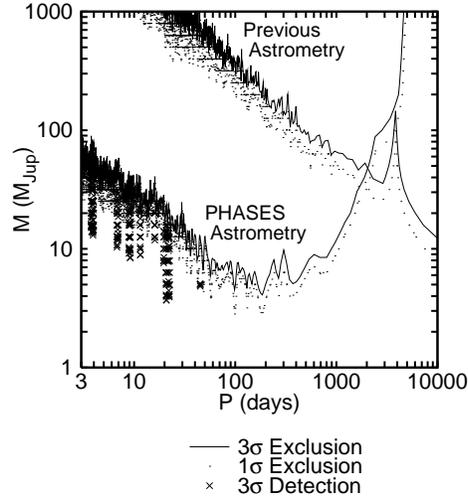}
\caption[$\beta$ Coronae Borealis Mass-Period Companion Phase Space] 
{ \label{fig:137909_phase_space}
The mass-period phase space for tertiary companions to $\beta$ Coronae 
Borealis shows that a proposed massive object with period $\sim 320$ days 
does not exist.  Some mass-period combinations do result in double-Keplerian 
fits that are an improvement over the single-Keplerian model, but the 
inner-body's orbit is poorly constrained and it is not clear whether these are 
astrophysical in origin.
}
\end{figure}

The excess multiplicative factor is 
$\chi^2_{\mathcal{A}} / Z_{\mathcal{A}} = 9.9 \sim 3.2^2$ 
for the PHASES observations.  The $1\sigma$ and $3\sigma$ exclusion 
contours for both the PHASES and non-PHASES data sets are shown in 
figure \ref{fig:137909_phase_space}.  In both cases, the $P_2 \sim 320$ day 
perturbation is excluded; for the PHASES observations, a companion of mass as 
small as $4\Mjup$ are excluded.  The mass limits should be increased 
by $\sim 1.77/1.21 \sim 1.46$ when considering the primary star as 
the companion host (this estimate assumes the companion is an 
insignificant part of the total mass, reasonable in the 
case of the lower limit exclusion region).  It appears 
that the perturbation in \citeauthor{Neubauer1944}'s data set is 
not astrophysical in origin.

There are regions of the mass-period phase space for which perturbations 
from a tertiary companion improve the fit by $3\sigma$ or more.  The 
smallest value of $\chi^2_{\mathcal{B}}$ corresponds to a $5.2\sigma$ 
detection at $P_2 \sim 20.9$ days.  When the restriction to face-on, the 
narrow orbit's parameters are strongly covariant and poorly constrained:  for 
example, the formal uncertainty on the eccentricity is 0.5, 
implying it is not constrained at all, and the Eulerian angles all have formal 
uncertainties greater than 360 degrees.  While the double-Keplerian model can 
formally improve the fit compared to the single version, the double-Keplerian 
model is too detailed to be properly constrained by the current observations.  
In order to claim a true detection, multiple criteria must be met:  the 
double-Keplerian model must better fit the data, and the perturbation orbit 
must be well-constrained.  Thus, it is possible that the 
``detection'' periodicities are astrophysical in nature, but the data 
sampling in the present set prevents conclusive study, and aliasing 
causes multiple periodicities to exist.

\section{13 Pegasi}

13 Pegasi (HR 8344, HD 207652) has been observed 25 times over a span of 
$T \sim 441$ days in the PHASES program.  The span is short compared to 
the binary period \citep[$P_1 \sim 26.132$ yr; ][]{Hart1989} 
and again $P_1$, $e_1$, and $a_1$ 
are covariant; in the analysis $P_1$ is thus held fixed at this nominal 
value ($Q_{\mathcal{A}} = 6$, $Q_{\mathcal{B}} = 9$).  \cite{Soder1999} 
used {\em Hipparcos} measurements to determine the system parallax of 
$29.6 \pm 0.8$ mas, a total mass of $2.67\Msun$, and photometric mass 
ratio $M_{\rm {B}}/M_{\rm {A}} = 0.80$.  Computations assume a 
distance of 33.8 pc and the tertiary companion orbits the secondary at 
mass $1.19 \Msun$ (the results can be scaled appropriately for the primary 
at mass $1.48 \Msun$.

\begin{figure}[tbh]
\epsscale{0.4}
\plotone{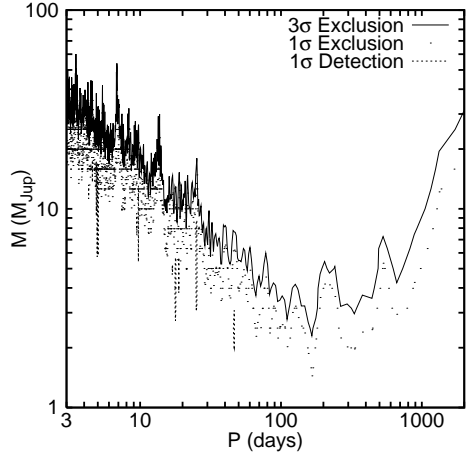}
\caption[13 Pegasi Mass-Period Companion Phase Space] 
{ \label{fig:207652_phase_space}
The 13 Pegasi Mass-Period companion phase space shows PHASES observations 
can rule out tertiary objects as small as two Jupiter masses.  A few 
mass-period combinations introduce slight improvements over the 
single-Keplerian model, but none of these are more significant than 
$1.7\sigma$, and are probably not astrophysical in origin.}
\end{figure}

The best fit $\chi^2_{\mathcal{A}} = 148.8$ with $Z_{\mathcal{A}} = 44$ 
degrees of freedom; assuming model $\mathcal{A}$ is correct, the 
multiplicative excess factor is $3.38 \sim 1.84^2$.  The median formal 
minor axis uncertainty of $10.2 \microas$ is thus adjusted to 
$18.8 \microas$ repeatability from night to night.  The mass-period grid 
is set up as in the case of $\delta$ Equulei.  Figure 
\ref{fig:207652_phase_space} shows that massive planets in 20 day to 3 
year period face-on circular orbits would perturb this binary by more 
than the observed scatter in the PHASES data.  Planets as small as 
two Jupiter masses are ruled out in $\sim 4$ month period orbits.  A 
number of regions show $1\sigma$ consistent detections, the best of 
these is at only the $1.7\sigma$ detection level.

\section{20 Persei}

The binary 20 Persei (HR 855, HD 17904, ADS 2200) has been observed 26 
times by PHASES.  At $K=4.3$ it is one of the fainter PHASES targets.  
The observations span $T \sim 877$ days, a small amount compared to the 
orbital period of $P_1 \sim 31.528$ yr.  Covariances between $P_1$, 
$e_1$, and $a_1$ are extremely strong.  Using values from 
\cite{Doc2001c}, both $P_1$ (31.528 yr) and $e_1$ ($0.753$) are 
fixed ($Q_{\mathcal{A}} = 5$, $Q_{\mathcal{B}} = 8$).  
The system distance is assumed to 
be 72.1 pc from the {\em Hipparcos} parallax; combining this 
with the orbital period and semimajor axis an average component 
mass of $2.1\Msun$ is computed and these values are 
used in converting tertiary mass to $a_2$.  The mass-period grid 
is set up following the same procedure as for $\delta$ Equulei.  For 20 Per, 
$\chi^2_{\mathcal{A}}/Z_{\mathcal{A}} = 6.57 \sim 2.56^2$ is a 
relatively small value.  However, the average formal minor-axis uncertainty 
is $21 \microas$; coupling this with the large distance and masses implies 
that the tertiary detection limits are not particularly sensitive.

\begin{figure}[tbh]
\epsscale{0.4}
\plotone{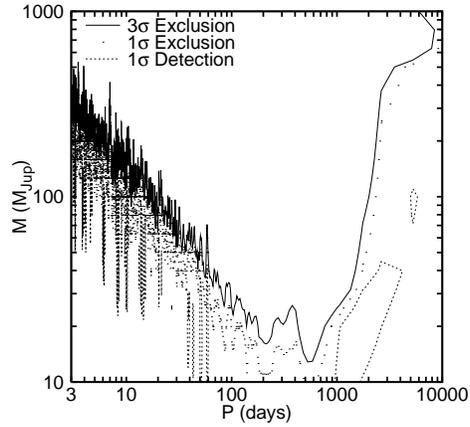}
\caption[20 Persei Mass-Period Companion Phase Space] 
{ \label{fig:17904_phase_space}
A range of brown dwarfs companions can be ruled out by the 20 Persei 
mass-period phase space.  Unfortunately, the PHASES observations do not yet 
have long enough time coverage to confirm the 1269 day period companion 
of \cite{abt1976}.  Owing to the far distance of the system and higher mass 
stars, the PHASES observations are not particularly sensitive to planetary 
companions in this system.}
\end{figure}

The PHASES measurements rule out brown dwarf mass companions in a variety of 
orbital periods, as shown in figure \ref{fig:17904_phase_space}.  Detections 
are found for several different companion periods at the $1\sigma$ detection 
level.  The smallest value of $\chi^2_{\mathcal{B}}$ appears at 
the $2.9\sigma$ level; as in the case of $\beta$ CrB, when the face-on and 
circular restrictions are lifted from the perturbation orbit, its parameters 
become poorly constrained.  \cite{abt1976} proposed the existence of a 
tertiary companion with period $P_2 = 1269 \pm 70$ days; this periodicity is 
very close to the steep cutoff in the phase space contour due to the finite 
span of PHASES observations; at this time no conclusions are made about the 
existence of such a companion.

\section{$\eta$ Coronae Borealis}

Seventeen PHASES observations of $\eta$ Coronae Borealis (2 CrB, HR 5727, 
HD 137107, ADS 9617) have been collected over a span of $T \sim 680$ days.  
As with 20 Per, $P_1$ and $e_1$ are fixed at values of 
$\sim 41.585$ yr and 0.2620 determined by speckle interferometry 
\citep{Msn1999a} to lift strong covariances with $a_1$.  
From {\em Hipparcos} data, \cite{Soder1999} determined a parallax of 
$53.5 \pm 0.9$ mas, total mass $2.41\Msun$, and photometric mass ratio 
of 0.96.  From these, the tertiary orbit to $a_2$ conversion is made assuming 
a distance of 18.69 pc and stellar mass of $1.18 \Msun$ (the primary's mass 
is roughly $1.23 \Msun$ and the tertiary companion limits are similar for 
primary and secondary).  The mass-period grid is set up as in the case of 
$\delta$ Equulei.  

Companions as small as 3-4$\Mjup$ in orbits of $\sim$100-700 days are 
inconsistent with the PHASES observations.  As in the case of $\beta$ CrB, 
detection contours deeper than $3\sigma$ appear (here at periods less than 
10 days).  The deepest of these is at the $4.8\sigma$ level.  
Similarly to $\beta$ CrB, the data sampling does not allow all parameters 
of the perturbation orbit to be well constrained when the restrictions of 
face-on and circular orbits are lifted (in particular, $e_2$ is not 
constrained at all).  This may be astrophysical in origin, but the current 
data cannot properly address the underlying orbit.

\begin{figure}[tbh]
\epsscale{0.4}
\plotone{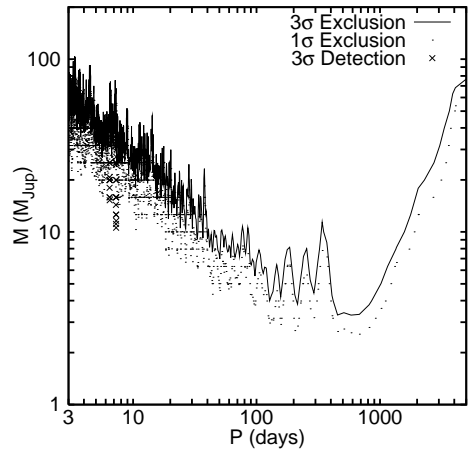}
\caption[$\eta$ Coronae Borealis Mass-Period Companion Phase Space] 
{ \label{fig:137107_phase_space}
The $\eta$ Coronae Borealis Mass-Period companion phase space shows planets 
as small as three Jupiter masses are inconsistent with PHASES observations.  
A nearly $5\sigma$ significant detection is found at short periods, but 
observational coverage does not yet allow the orbit to be strongly 
constrained; it is unclear whether this is merely a statistical fluke, perhaps 
related to the data sampling function (both dips are near 7 days, a natural 
sampling time for PHASES observations).}
\end{figure}

\section{HR 7162}

HR 7162 (HD 176051, ADS 11871) is a G0V primary with K1V secondary.  This 
color difference appears to affect the PHASES data through the differential 
dispersion mechanism because the excess scatter 
$\chi^2_{\mathcal{A}}/Z_{\mathcal{A}}$ is much larger than for other stars 
($\sim 88 \sim 9.4^2$).  As with 20 Per and $\eta$ CrB, both 
$P_1$ and $e_1$ are fixed at values of $\sim 61.15$ yr and 
0.258 determined by \cite{Hei1994a} to lift strong 
covariances with $a_1$.  Over a span of $\sim 773$ 
days, 32 PHASES observations have been collected.  The tertiary mass 
to $a_2$ conversion assumes a distance of 15 parsecs and star mass 
$0.71 \Msun$.  These quantities are derived from the {\em Hipparcos}-based 
solution of \cite{Soder1999}, who found a parallax of $66.7 \pm 0.6$ mas, 
mass sum $1.78\Msun$ and photometric mass ratio 0.67 (the primary mass 
corresponds to $1.07\Msun$).  The mass-period grid is constructed as in 
the case of $\delta$ Equulei.  The smallest values of $\chi^2_{\mathcal{B}}$ 
correspond to a $2.7\sigma$ detection.  Companions as small as $2\Mjup$ are 
inconsistent with the PHASES data.

The physical properties of the HR 7162 binary (low mass stars, 
close to Earth) make it an exciting candidate 
for an exoplanet search as very low mass companions can be detected for a 
given astrometric performance levels.  Unfortunately, the PHASES excess 
scatter factor for this binary is currently much larger than that typical in 
other systems studied, and appears to be random noise.  This is likely 
due to the large color difference of the binary components, potentially 
leading to differential dispersion issues.  It is anticipated that the current 
engineering improvements will benefit this system in particular, in which case 
it will be an extremely exciting study for the exoplanet search, with the 
ability to detect (or rule out) planets as small as a fifth of a Jupiter 
mass orbiting either star.  

\begin{figure}[tbh]
\epsscale{0.4}
\plotone{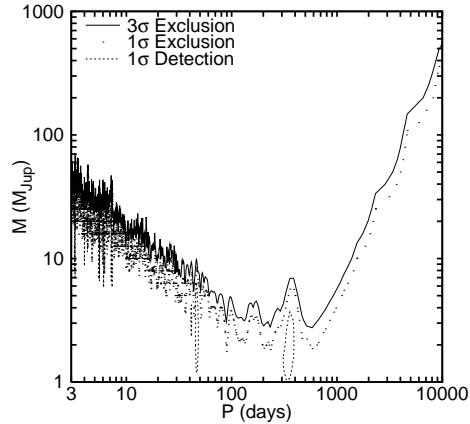}
\caption[HR 7162 Mass-Period Companion Phase Space] 
{ \label{fig:176051_phase_space}
Contours in the companion mass-period phase space for HR 7162 show that 
despite high noise levels from systematic processes, companions as small as 
two Jupiter masses can be excluded in face-on orbits.  After engineering 
improvements to remove these systematic noise sources, constraints will be 
placed on sub-Jupiter massed planets by second stage PHASES observations.
}
\end{figure}

\section{Conclusions}

PHASES observations are able to exclude tertiary companions with masses as 
small as a few Jupiter masses in several binary systems.  The phase-space 
searching algorithm correctly identifies the natures of two previously studied 
triple star systems and finds the correct periods for the short period pairs.  
While the number of systems that can currently be examined is too small to 
make strong conclusions about the frequency of giant planets in close binary 
systems, by the end of the PHASES program enough measurements will be 
made on all target systems to address this important question.  Systematic 
effects currently limit the observed precision; if overcome, companions as 
small as a fifth of a Jupiter mass might be detected in favorable systems 
such as HR 7162.

\acknowledgements 

PHASES benefits from the efforts of the PTI collaboration members who have 
each contributed to the development of an extremely reliable observational 
instrument.  Without this outstanding engineering effort to produce a solid 
foundation, advanced phase-referencing techniques would not have been 
possible.  We thank PTI's night assistant Kevin Rykoski for his efforts to 
maintain PTI in excellent condition and operating PTI in phase-referencing 
mode every week.  We thank Dimitri Pourbaix for helpful conversations about 
astrometric orbit fitting.  
Part of the work described in this paper was performed at 
the Jet Propulsion Laboratory under contract with the National Aeronautics 
and Space Administration. Interferometer data was obtained at the Palomar
Observatory using the NASA Palomar Testbed Interferometer, supported
by NASA contracts to the Jet Propulsion Laboratory.  This research has made 
use of the Washington Double Star Catalog maintained at the U.S.~Naval 
Observatory.  This research has made use of the Simbad database, operated 
at CDS, Strasbourg, France.  MWM acknowledges the MIT Earth, Atmospheric, 
and Planetary Sciences department for hosting him while this paper was 
written.  BFL acknowledges support from a Pappalardo Fellowship in Physics.  
PHASES is funded in part by the California Institute of Technology Astronomy 
Department, and by the National Aeronautics and Space Administration under 
Grant No.~NNG05GJ58G issued through the Terrestrial Planet Finder Foundation 
Science Program.  This work was supported in part by the National Science 
Foundation through grants AST 0300096 and AST 0507590.

\bibliography{main}
\bibliographystyle{plainnat}

\end{document}